\documentstyle[12pt,epsf,psfig,bm]{article}
\begin{document}

\begin{center}
{\Large\bf Effects of the crystal structure in the dynamical electron 
density-response of hcp transition metals} \\

\vspace{0.5cm}
{ I. G. Gurtubay,$^{\rm (1)}$ 
Wei Ku,$^{\rm (2)}$ 
J. M. Pitarke,$^{\rm (1,3)}$ 
and 
A. G. Eguiluz$^{\rm (4,5)}$} \\

\vspace{0.6cm}
{\it $^{\rm (1)}$ Materia Kondentsatuaren Fisika Saila, Zientzi Fakultatea,
Euskal Herriko Unibertsitatea, 644 Posta kutxatila, 48080 Bilbo, Basque Country,
Spain}\\
{\it $^{\rm (2)}$ Department of Physics, Univ. of California, Davis, CA 95616-8677, USA}\\
{\it $^{\rm (3)}$ Donostia International Physics Center (DIPC) and Centro Mixto
CSIC-UPV/EHU, Donostia, Basque Country, Spain}\\
{\it $^{\rm (4)}$ Department of Physics and Astronomy, Univ. of Tennessee, 
Knoxville,
TN 37996-1200, USA}\\
{\it $^{\rm (5)}$  Solid State Division, Oak Ridge National Laboratory, 
Oak Ridge, TN 37831-1200, USA}\\
\end{center}

\begin{abstract}
We present an all-electron study of the dynamical density-response function of
hexagonal close-packed transition metals Sc and Ti.  We elucidate various
aspects of the interplay between the crystal structure and the electron
dynamics by investigating the loss function, and the associated dielectric
function, for wave-vector transfers perpendicular and parallel to the
hexagonal plane. 
As expected, but contrary to recent work, we find that the
free-electron-like aspects of the dynamical response are rather isotropic
for small wave vectors.
The crystal local-field effects are found to have an impact on the 
plasmon energy for {\it small wave vectors}, 
 which gives rise to an interplay with the exchange-correlation 
effects built into the many-body kernel. 
The loss function lineshape shows a significant dependence on 
propagation direction; in particular, for propagation on the hexagonal 
plane the plasmon hybridizes substantially with fine structure 
due to $d$-electron transitions, and its dispersion curve becomes difficult 
to establish, beyond the small wave vector limit. 
The response is calculated in the framework of time-dependent
density functional theory (TDDFT),
based on a full-potential 
linearized augmented-plane-wave (LAPW) ground-state, in which 
the exchange-correlation effects are treated in the local-density 
approximation. 
\end{abstract}

\vspace{0.2cm}
Electronic excitations in solids can be studied from the knowledge of the 
dynamical density-response function, which is directly related to observables 
such as inelastic scattering cross sections 
for fast electrons and  hard x-rays.

For many years, theoretical investigations of valence-electron excitations
were carried out on the basis of the  
free-electron gas (FEG) model (or jellium model), 
in which valence electrons are described by an
assembly of interacting electrons embedded in a uniform 
background of positive charge.
The physics of the FEG model is completely controlled by only one parameter,
 the valence-electron density,
$n_0$ or, equivalently, 
by the electron-density parameter $r_s$ which is related 
to $n_0$ through $1/n_0=4\pi/3(r_s a_0)^3$,
$a_0$ being the Bohr radius.
As we shall illustrate below, in real materials the
 physics of the excitations is more complex; indeed, the loss spectra 
shows qualitative departures from the FEG description.

In the past few years,
{\it ab initio} calculations of the density-response function 
of many simple
\cite{AL}-\cite{EKS} and noble \cite{Cu-Igor}-\cite{Au}
 metals have been reported.
Interestingly, in a number of cases it has been concluded
 that  remarkable anomalies revealed by 
experimental measurements arise mainly from band-structure effects 
\cite{examples}.  Examples include 
 the fine structure in the inelastic X-ray 
scattering data for aluminum
\cite{AL}, the negative plasmon energy
dispersion of cesium \cite{CS},\cite{CS-Fleszar}, 
and the positive dispersion of the linewidth of the plasmon in potassium
\cite{K-th}  .

In this paper we utilize an all-electron  method to compute the 
density-response function
 of Sc and Ti starting from the knowledge of an LAPW ground state 
\cite{K-th},\cite{EKS},\cite{Zn}. 
Our presentation centers on a discussion of the dominant 
feature of the excitation spectrum for these 
hexagonal closed-packed transition metals for relatively small
 wave vectors --- the collective, plasmon-like mode, and its interplay with the
 effects of the crystal lattice; in particular, we assess the
impact of the so-called crystal local fields, 
and the dependence on propagation direction.  
We conclude that the crystal local-field effects 
are not negligible, in the small wave vector limit;
this brings about an interplay with the effects of exchange and correlation,
which thus impact the plasmon dispersion for small wave vectors,
unlike the case of the FEG.
Similarly, the dependence on propagation direction of the response function
 is rather significant: for wave vector transfers along the hexagonal plane, 
the plasmon hybidizes with the fine structure caused by $d$-electron 
transitions and its dispersion curve becomes difficult to establish 
beyond the small wave vector limit.
Comparison is made with the recent pseudopotential-based calculations of 
the density-response function of Sc by
Sch\"{o}ne and Ekardt \cite{schone}; the surprising result
 reported by these authors 
 that the plasmon-like excitation is not realized for 
small wave vectors for propagation in the hexagonal plane 
 is not supported by our calculations. 
Our results agree with the intuitive expectation 
that the free-electron-like response is not affected drastically 
by the propagation direction.

\section{Theory}
The density-response function $\chi({\bf r},{\bf  r'};\omega)$
\cite{P-Noz} of an interacting electron system 
gives, within linear-response theory,
 the electron
density induced in the system when the latter is exposed 
to an external potential $V^{ext}({\bf r},\omega)$  
through the equation
\begin{eqnarray}\label{eq1}
\rho^{ind}({\bf r},\omega)=\int{\rm d}{\bf r}'\,
\chi({\bf r},{\bf  r'};\omega)\,V^{ext}({\bf r'},\omega).
\end{eqnarray}

 In the framework of 
 time-dependent density-functional theory (TDDFT) \cite{Runge-Gross}, 
the $exact$ density-reponse function
$\chi({\bf r},{\bf  r'};\omega)$ obeys   the integral equation 
\cite{Peter} 
\begin{eqnarray}\label{eq2}
\chi({\bf r},{\bf r}';\omega)=\chi^S({\bf r},{\bf r}';\omega) 
+\int{\rm d}{\bf r}_1\int{\rm d}{\bf r}_2\chi^S({\bf r},{\bf r}_1;\omega)
\cr\cr
 \times 
\left[ v({\bf r}_1-{\bf r}_2)+f_{xc}({\bf r}_1,{\bf r}_2;\omega)
\right]
\chi({\bf r}_2,{\bf r}';\omega),
\end{eqnarray}
where $\chi^S({\bf r},{\bf r}';\omega) $ is the single-particle 
density-response function of the $unperturbed$ Kohn-Sham (KS) system
\cite{Kohn}, 
$v({\bf r}-{\bf r}')$ is the bare Coulomb interaction, 
and $f_{xc}({\bf r},{\bf r'};\omega)$ accounts for all dynamical
exchange-correlation (xc) effects.

For a periodic crystal it is  convenient to work in Fourier
space, so we write 
\begin{equation}\label{eq3}
\chi({\bf r},{\bf r}',\omega)={1\over\Omega}\sum_{\bf k}^{BZ}\sum_{{\bf
G},{\bf G}'}{\rm e}^{{\rm i}({\bf k}+{\bf G})\cdot{\bf r}}{\rm
e}^{-{\rm i}({\bf k}+{\bf G}')\cdot{\bf r}'}\chi_{{\bf G},{\bf G}'}({\bf
q},\omega),
\end{equation}
 with a similar Fourier expansion for  $\chi^{S}({\bf r},{\bf  r'};\omega)$.
The Fourier  coefficients of the KS response function,
$\chi^{S}_{{\bf G},{\bf G}'}({\bf k},\omega) $,
can be written as
\begin{eqnarray}\label{eq4}
\chi_{{\bf G},{\bf G}'}^S({\bf k},\omega)={1\over \Omega}\sum_{\bf
k'}^{BZ}\sum_{n,n'} {f_{{\bf k'},n}-f_{{\bf k'}+{\bf k},n'}\over E_{{\bf
k'},n}-E_{{\bf k'}+{\bf k},n'} +\hbar(\omega + {\rm i}\eta)}\cr\cr
\times\langle\phi_{{\bf k'},n}|e^{-{\rm i}({\bf k}+{\bf G})\cdot{\bf
r}}|\phi_{{\bf k'}+{\bf k},n'}\rangle
\langle\phi_{{\bf k'}+{\bf k},n'}|e^{{\rm i}({\bf k}+{\bf G}')\cdot{\bf
r}}|\phi_{{\bf k'},n}\rangle,
\end{eqnarray}
where $\Omega$ represents the normalization volume, 
{\bf G} and {\bf G}' are vectors of the reciprocal lattice, 
and the second sum runs over the band structure for each wave vector
${\bf k'}$ of the first Brilloun Zone (BZ).
$\phi_{{\bf k},n}({\bf r})$ and
$E_{{\bf k},n}$ are  Bloch eigenfunctions and eigenvalues of the
KS Hamiltonian of ground-state density-functional theory (DFT) \cite{Kohn},
and $f_{{\bf k},n}$ are the occupation numbers, with due account for
spin degeneracy.  

In the above Fourier representation, eq. (\ref {eq2}) turns into a 
matrix equation 
\begin{eqnarray}\label{eq5}
\chi_{{\bf G},{\bf G}'}({\bf k},\omega)=\chi^S_{{\bf G},{\bf G}'}({\bf
k},\omega)+\sum_{{\bf G}''}\sum_{{\bf G}'''}\chi^S_{{\bf G},{\bf G}''}({\bf
k},\omega)\cr\cr
\times\left[v_{{\bf G}''}({\bf k})\delta_{{\bf G}'',{\bf G}'''}+f^{xc}_{{\bf
G}'',{\bf G}'''}({\bf k},\omega)\right]
\chi_{{\bf G}''',{\bf G}'}({\bf k},\omega),
\end{eqnarray}
which we solve  numerically. 
The size of this matrix equation is  a measure of the importance
of the crystal local-field effects, arising from the inhomogeneity 
of the electronic environment in the periodic crystal potential.
For later reference, we note that ignoring the crystal 
local-field effects corresponds to  solving eq. (\ref {eq5})  while
keeping only the 
zero-zero element of the KS  response matrix.

Within the first Born approximation   the inelastic
scattering cross-section for hard  x-rays 
and   fast electrons corresponding to wave-vector 
 transfer   {\bf q} = ${\bf k}+{\bf G}$, is 
proportional to the dynamic-structure factor
\begin{equation}\label{sqw}
S({\bf q}, \omega)= -2 \;\hbar \; \Omega\; {1 \over {v_{\bf G}({\bf k})} }
{\rm Im}\left[{\epsilon_{{\bf G},{\bf G}}^{-1}({\bf k},\omega)}\right],
\end{equation}
where  Im $ \left[ \rm\epsilon^{-1}_{{\bf G},{\bf G}'}({\bf k},\omega)\right]$
is the so-called energy-loss function and is related to the response 
function through
%
%
%
\begin{equation}
\epsilon_{{\bf G},{\bf G}'}^{-1}({\bf k},\omega)=\delta_{{\bf G},{\bf G}'}+v_{{
\bf
G}}({\bf k})\chi_{{\bf G},{\bf G}'}({\bf k},\omega).
\end{equation}
The plasmon-energy dispersion is given by the frequencies at which
the real part of the dielectric function,
$\epsilon({\bf k},\omega)$,
is close to zero and the imaginary part is small. Im$ \rm\epsilon^{-1}({\bf k},\omega)$ 
is therefore a maximum.
 Throughout this work  the plasmon-energy is taken to be  the
energy location of the main peak of the energy-loss function. 
\section{Numerical Implementation}
The key ingredient in the calculation of the energy-loss spectrum is 
the KS response function $\chi^S$ of eq. (\ref{eq4}). 
For this purpose, we need  the KS states which we  obtain  
 within the local-density approximation (LDA) \cite{PWang}, 
in terms of a  variational expansion  
  in the
LAPW basis \cite{lapw}:
\begin{eqnarray}\label{eq-lapw}
 \phi_{{\bf k},n}({\bf r})= 
\frac{1}{\Omega} \sum_{{\bf G}} C_n({\bf k}+{\bf G}) 
	\psi^{LAPW}_{{\bf k}+{\bf G}}({\bf r}).
\end{eqnarray}
The LAPW functions $\psi^{LAPW}_{{\bf k}+{\bf G}}({\bf r})$ are obtained
dividing the unit cell   into two regions: 
non-overlapping atomic spheres centered at nuclear sites 
and the interstitial region between the spheres. 
This allows a faithful description of the localized strong 
oscillations near the atomic site, where atomic-like functions are used,
and the smooth behaviour of the interstitial region, where 
plane waves are employed.
Local orbitals are also introduced in order to have an accurate description
of the semi-core states.

The use of the symmetry properties of the crystal allows us to work
with Bloch states that involve momentum transfers in the irreducible
Brillouin Zone (IBZ) instead of those in the whole zone.
If $\tilde{\rm \bf k}$ is a 
vector of the IBZ,
\begin{eqnarray}\label{Blochk}
 \phi_{{\bf k},\: n}({\bf r}) =  \phi_{{\rm R^{-1}{\bf \tilde{k}}},\: n}({\bf r}) = 
e^{-{\rm i }{\bf \tilde{k}}\,\cdot \, {\bm {\tau}}\rm(R) }
 \phi_{{\bf \tilde{k}},\: n}({\rm R} \cdot{\bf r} + {\bm \tau}) ,
\end{eqnarray}
where R(${\tilde{\bf k}}$) runs over the sub-set of the point group 
 of the crystal  that generates
the star of ${\bf \tilde{k}}$ and ${\bm \tau}$(R) is the partial lattice 
displacement corresponding to the space-group operation 
\{R, ${\bm \tau}$(R)\}.

The evaluation of the matrix elements in eq. (\ref{eq4})
is performed with a code that  runs 
in parallel and scales linearly with the number of nodes used. 
With the matrix elements  at hand we perform the sum  over  k-points and 
over the  pairs of bands allowed  by the occupation numbers.
We  solve  equation (\ref{eq5}) 
using two different approximations for $f_{xc}$. 
In the random-phase approximation (RPA),  $f_{xc}$ = 0. In 
the so-called adiabatic local-density approximation (ALDA),
\begin{equation}\label{eq7}
f_{\rm xc}^{\rm ALDA}({\bf r},{\bf r}';\omega)
=\delta({\bf r}-{\bf r}')\left [{{\rm d}^2
E_{\rm xc}(n)\over {\rm d}n^2} \right ]_{n=n_0({\bf r})},
\end{equation}
where $E_{\rm xc}(n)$ is the exchange-correlation energy of a 
uniform electron gas of density $n$, and $n_0({\bf r})$
is the ground-state density.

We apply the above     formalism to
hexagonal closed-packed transition metals Sc and Ti, whose electronic 
configurations  are [Ar]3$d^{1}$4$s^{2}$ and [Ar]3$d^{2}$4$s^{2}$,
 respectively.
The lattice parameters used are 
$ a $=2.95 $\rm\AA$ and $ c$=4.68 $\rm\AA$ for Ti and 
$ a $=3.31 $\rm\AA$ and $ c$=5.28 $\rm\AA$ for Sc \cite{Ashcroft}.
The ground state is obtained with a 
 cut-off parameter
$R_{MT}K_{max}$=8.
Inside the atomic spheres, the LAPW wave functions are expanded
in spherical harmonics $Y_{lm}$ up to $l$=10.
The response function was evaluated  for {\bf q}
wave vectors
perpendicular and parallel to the hexagonal plane, using 8 x 8 x 16 and 
16 x 16 x 8 BZ samplings (the corresponding number of points in the IBZ 
being 90 and 150, respectively) 
and keeping KS states up to 7.5Ry, thereby ensuring convergence 
in the energy range under study.
The 3$s$ and 3$p$ states were treated as semi-core states.
\section{Results}
\label{Results}
%
%
\begin{figure}[t]
\epsfclipon
\epsfxsize=8cm 
\centerline{\psfig{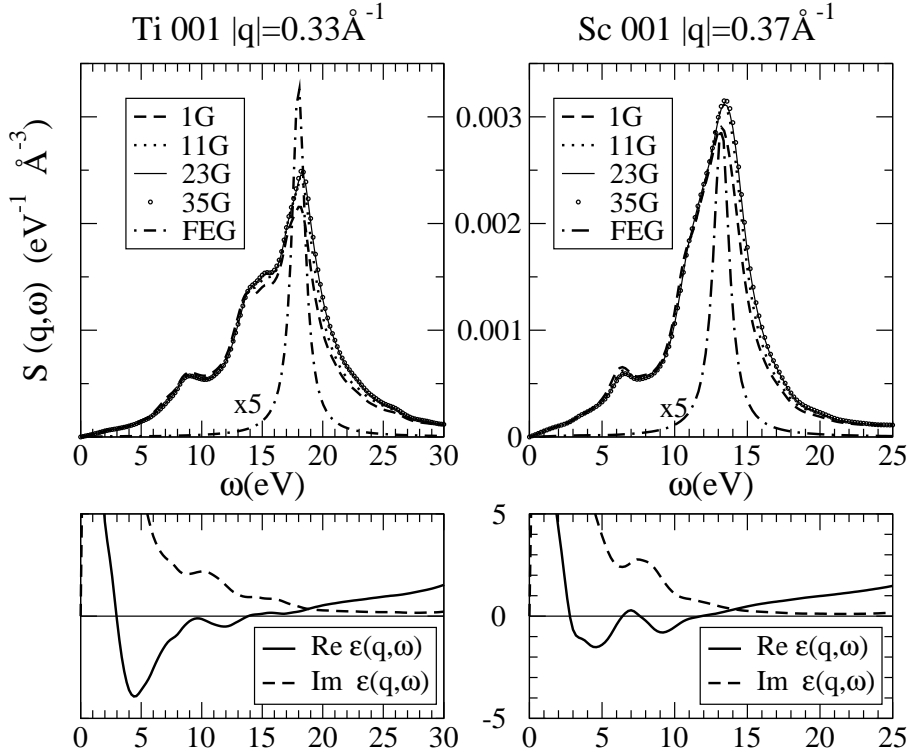}}
\caption{\protect\small 
Top panel:
impact of local-filed effects on the dynamic structure function
 of Ti and Sc along the hexagonal
axis within ALDA. The values of the momentum transfer are 
{\bf q}= 4/16(001) and {\bf q}= 5/16(001) respectively, both in units
of 2$\pi/c$.
The dotted-dashed line corresponds to a FEG calculation
with $r_s$=1.92 a.u. (Ti) and $r_s$=2.38  a.u. (Sc). 
Lower panel: dielectric function.
}
\label{fig:LFE}
\end{figure}
%
\begin{figure}[t]
\epsfclipon
\epsfxsize=8cm 
\centerline{\psfig{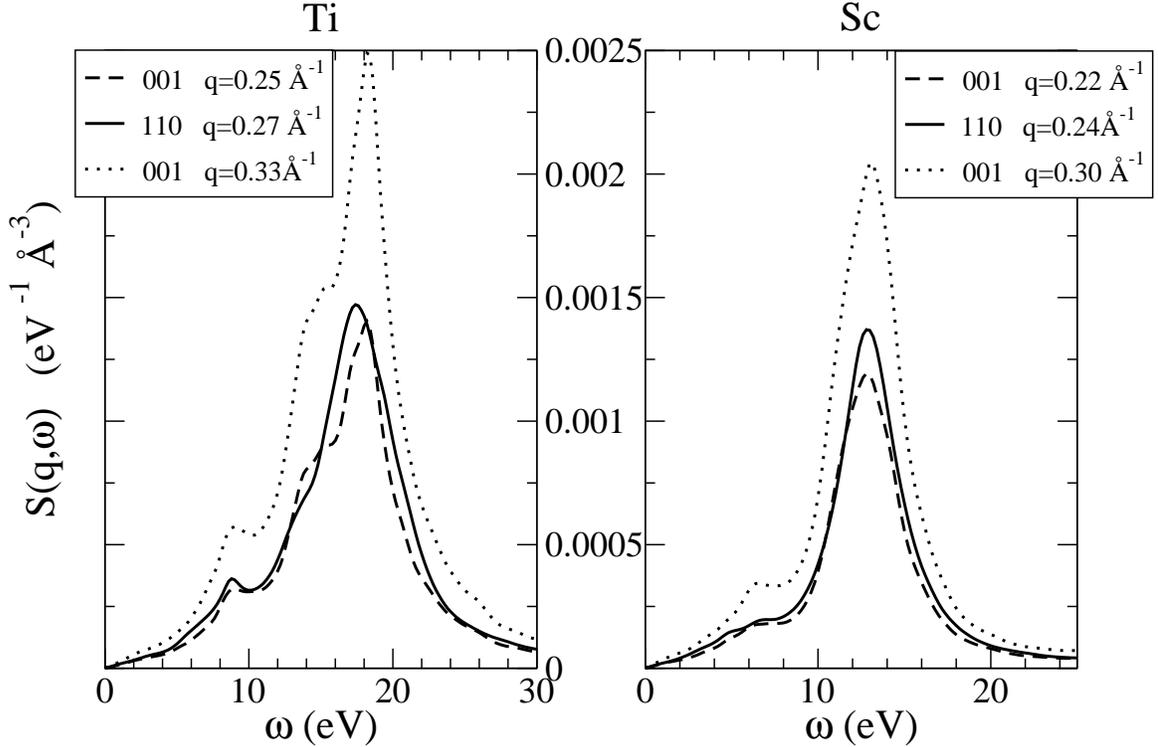}}
\caption{\protect\small 
Comparison of the dynamical structure factor 
for similar values of the momentum transfer perpendicular (001)
and parallel to the hexagonal plane (110) for Ti (left) and Sc (right)
 within ALDA.
}
\label{fig:isotr}
\end{figure}
%
\begin{figure}[t]
\epsfclipon
\epsfxsize=8cm
\centerline{\psfig{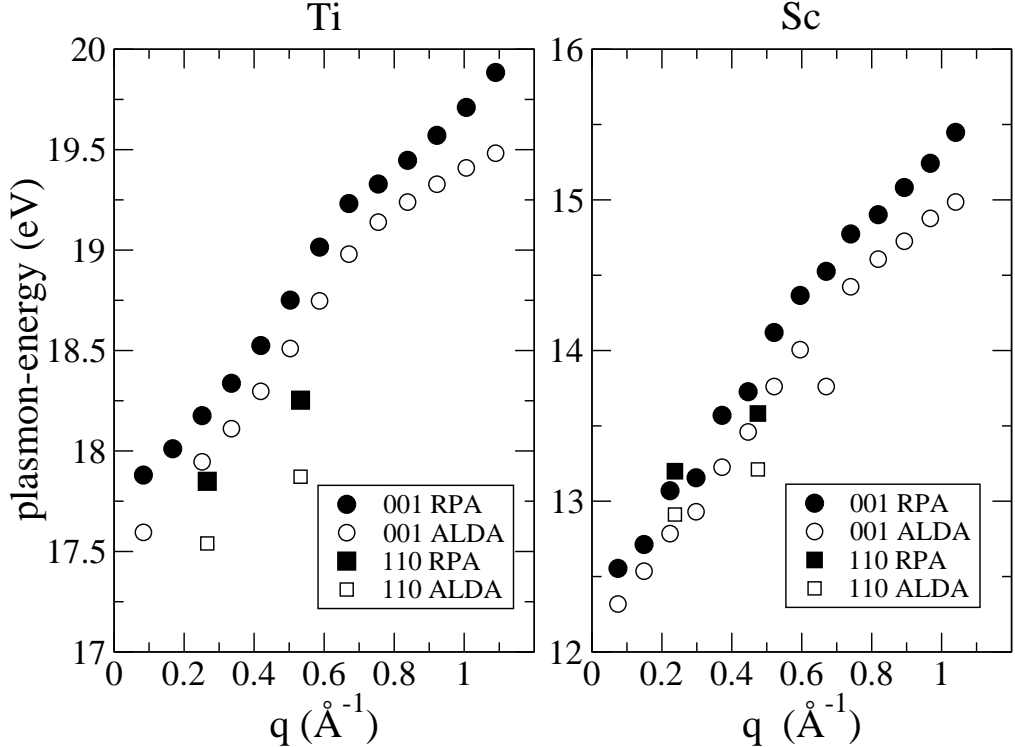}}
\caption{\protect\small 
Plasmon-energy dispersion for Ti (left) and Sc (right) along the 001
 direction (circles) and along the 110 direction (squares)
within RPA (solid symbols) and within ALDA (empty symbols). 
}
\label{fig:plas}
\end{figure}
%
The upper  left (right)   panel in Fig. \ref{fig:LFE} shows the 
dynamical-structure factor obtained within ALDA
 for Ti (Sc) for energies up to 
the onset of semicore excitations 
($\sim$ 32.5 eV in Ti and $\sim$ 27.7 eV in Sc)
 along the (001) direction.
We have inverted eq. (\ref{eq5}) using  up to 9 shells (55 {\bf G}-vectors)
 \cite{Gvec}. 
For clarity, only results for the scalar inversion, corresponding
to a  1x1 matrix calculation with the 
{\bf G} = {\bf G}$'$ = 0 element (dashed line), and for matrix inversion
using 11 (dotted-line), 23 (solid line) and 35  (open circles) 
{\bf G}-vectors  are displayed.
It's apparent that convergence is reached for the energy 
range under study
 (the physics of the dynamical response above the M-edge 
will be the subject of another publication); 
the results shown in Figs. \ref{fig:isotr} and \ref{fig:plas}
correspond to the converged  35 {\bf G}-vector calculations
   \cite{eta1}.
 $S({\bf q}, \omega$) displays a broad plasmon-like 
peak,  whose
position is shifted  upwards (between 0.5 eV and 1.0 eV) by the crystal
 local-field effects. 
These peaks correspond
(see lower panels of Fig. \ref{fig:LFE}) to zero values of 
Re $\epsilon({\bf q},\omega)$ 
where Im $\epsilon({\bf q},\omega)$ is small.

For comparison, the upper panels of Fig. \ref{fig:LFE} also show
 (dotted-dashed line) the dynamical structure factor calculated within  the 
FEG model for $r_s$=1.92 a.u. (Ti) and $r_s$=2.38 a.u. (Sc) for
the same wave-vector
as the {\it ab initio} calculation, 
$r_s$ being the electron density 
equivalent to that of valence electrons, 3$d^{1}$4$s^{2}$ and 3$d^{2}$4$s^{2}$,
in Ti and Sc respectively \cite{etafeg}.
Clearly, as anticipated at the outset, the overall loss functions of 
these transition metals bear little resemblance to the FEG model, 
other than the presence of a broad collective mode; 
again, the essentially overdamped mode which defines the leading 
feature in Fig. \ref{fig:LFE} corresponds to a dielectric function whose nature 
is quite removed from the simple, clean, Drude-like zero which defines 
the textbook plasmon. 
Additional peaks  around 9 eV in Ti and around 6 eV in 
Sc arise from transitions involving  the $d$-electrons  
and will be discussed elsewhere.

In order to investigate the dependence on the direction of 
the momentum transfer, we show in
Fig. \ref{fig:isotr}  the ALDA dynamic structure factor S({\bf q}, $\omega$)
for Ti and Sc along the (001) and (110) directions for q values of the 
same order of magnitude.
Dashed and dotted lines represent  $S({\bf q}, \omega)$ for
{\bf q}= 3/16(001) and {\bf q}= 4/16(001), respectively, in units of 
2$\pi/c$. 
The calculation along the (110) direction (solid line) 
is for {\bf q}= 1/16(110), in units
of 2$\pi/a$. 
The curves along the (001) direction are calculated with a 8 x 8 x 16 
sampling over the BZ and the ones in the (110) direction with a
16 x 16 x 8 sampling
 \cite{sam-eta}. 
As the left panel on Fig.  \ref{fig:isotr} illustrates, 
for both propagation directions the plasmon peak in Ti
hybridizes with  fine structure due to $d$-electron transitions 
(prominent peak and shoulder at its left, respectively). 
This hybridization is more pronounced along the hexagonal plane. 
 For this reason, 
 the identification of a plasmon peak for
larger momentum transfers along this direction becomes very ambiguous.

In  recent work \cite{schone}, Sch\"{o}ne and Ekardt have 
reported pseudopotential-based calculations of the dynamic
structure factor  for Sc.
  The intriguing aspect of their results is that they suggest 
that the collective mode is not realized for small wave vectors for 
propagation along the hexagonal plane.  
By contrast, as shown in Fig. \ref{fig:isotr} the nature of the 
collective-like response in Sc (and also in Ti) is, for small wave vectors,
  quite similar for propagation directions parallel and perpendicular 
to the hexagonal plane.  
Our result agrees with intuitive expectations; on that basis we submit that it is correct.

Fig. \ref{fig:plas} shows the plasmon-energy dispersion for both  directions
within RPA (solid symbols) and ALDA (empty symbols).
Circles and squares correspond to wave-vector transfers 
that are perpendicular and parallel to the 
hexagonal plane.
As expected, the  ALDA 
plasmon energy shifts to slightly smaller values compared
to that of the RPA, due to a less effective electron-electron interaction.

An interesting feature in both panels of Fig. \ref{fig:plas} is that, 
in the small 
wave vector limit, RPA and ALDA do not yield the same result, as it occurs
in the FEG. Indeed, in the case of the FEG, the Coulomb term dominates
over the exchange-correlation contribution when {\bf q} $\rightarrow$ 0.
In the case of the {\it ab initio} calculation,
the 
difference  between RPA and ALDA 
arises from the inclusion of   crystal local-field
effects and its interplay with the exchange correlation effects: 
for small wave vectors, {\bf q}
 in eq. (\ref{sqw}) lies in the BZ and hence, 
the element  we consider after solving the matrix equation (\ref{eq5})
for $\chi_{{\bf G},{\bf G'}}$
is the {\bf G} = {\bf G}'= 0 element;
thus, in this case {\bf k} = {\bf q}. 

Through the inversion process there is
a  ``mixing'' between large wave vector arguments {\bf k}+{\bf G} in
$\chi^S$ and the small wave vector {\bf q} of the plasmon,
causing  a shift of the collective peak to larger energies. 
On the other hand, as mentioned before, 
 exchange-correlation effects
cause a downwards shift, smaller than the former one, provoking thus
the difference \cite{CLFq0}. The partial cancellation between both effects
leads to a net difference between the converged, ALDA result, and the 
scalar RPA result, for the small-q plasmon dispersion curve.

Ti and Sc show positive plasmon dispersions  and 
 disperse faster  along the
hexagonal axis.
For $|{\bf q}|$ values larger than $\sim \rm {0.5 \AA^{-1}}$, the assignment
of a collective peak in the (110) direction becomes a difficult task.
By contrast, this assignment is still possible along the (001) direction.
This fact ratifies, once more, the importance of the underlying crystal 
structure.

 We note that for Ti, the plasmon energy ($\sim$ 19 eV)  is in good agreement with the one 
measured in reference \cite{Macrander} for $q$=0.7 $\rm\AA^{-1}$ along (001).

\section{Conclusions}
We have presented results of calculations of 
 the dynamic response function of hcp Sc and Ti based on a
full-potential linearized augmented plane wave method. 
We have found that crystal local-field effects have a non-negligible impact
on the plasmon energy for small wave vectors causing an interplay 
between the crystal potential and the exchange-correlation effects.
We have investigated the dependence on propagation  direction of the
dynamical-structure factor and we have shown that, contrary to a recent 
report \cite{schone},
the plasmon loss  is quite  isotropic; 
however, the lineshape of the energy loss shows considerable dependence on the 
direction of the wave vector transfer: along the hexagonal plane, the 
identification of a dispersion relation for the plasmon energy
beyond the small wave vector limit,
becomes more difficult due to hybridization with 
the fine structure originated by $d$-electron transitions.

I. G. G. acknowlegdes support by the Basque Government, through the 
Hezkuntza, Unibertsitate eta Ikerketa Saila. I. G. G. is grateful to the 
hospitality of the University of Tennessee, Knoxville and the Solid State
Division at Oak Ridge National Laboratory, where most of this work has been 
carried out. A. G. E. acknowledges support from NSF ITR DMR-0219332.


\begin{thebibliography}{99}
\bibitem{eta1}
The numerical damping introduced in eq. (\ref{eq4}), $\eta$ ,
  has been set equal to 0.65 eV. 
Smaller values of $\eta$ sharpen the structures of the energy-loss
function without adding new information.
\bibitem{etafeg}
The numerical broadening  used is also 0.65 eV and is responsible for
the finite linewidth of the FEG peak.
\bibitem{sam-eta}
This choice   of the samplings over the BZ  for each direction
allows us to use the same value for $\eta$ and therefore the
comparison of the lineshapes in both directions is meaningful.
\end{thebibliography}
\end{document}